\documentclass[article,twocolumn,
superscriptaddress,
amsmath,amssymb,
aps,
pra,
floatfix,
]{revtex4-2}

\usepackage[normalem]{ulem}
\usepackage{graphicx}
\usepackage{xfrac}
\usepackage{amsmath}
\usepackage{mathtools}

\usepackage{epsfig}
\usepackage[colorlinks=true,citecolor=blue,linkcolor=blue,urlcolor=blue]{hyperref}
\usepackage{MnSymbol}
\usepackage{siunitx}
\usepackage{float}

\usepackage{graphicx}
\usepackage{dcolumn}
\usepackage{bm}
\usepackage{tikz,xcolor,hyperref}

\definecolor{lime}{HTML}{A6CE39}
\DeclareRobustCommand{\orcidicon}{
	\begin{tikzpicture}
	\draw[lime, fill=lime] (0,0) 
	circle [radius=0.16] 
	node[white] {{\fontfamily{qag}\selectfont \tiny ID}};
	\draw[white, fill=white] (-0.0625,0.095) 
	circle [radius=0.007];
	\end{tikzpicture}
	\hspace{-2mm}
}
\foreach \x in {A, ..., Z}{\expandafter\xdef\csname orcid\x\endcsname{\noexpand\href{https://orcid.org/\csname orcidauthor\x\endcsname}
			{\noexpand\orcidicon}}
}

\begin{document}

\title{Study of the elusive $5s-4f$ level crossing in highly charged osmium with optical transitions suitable for physics beyond the Standard Model searches}
    
\author{Nils-Holger Rehbehn\orcidA{}}
\affiliation{Max-Planck-Institut f\"ur Kernphysik, D--69117 Heidelberg, Germany}

\author{Lakshmi Priya Kozhiparambil Sajith}
\affiliation{Max-Planck-Institut f\"ur Kernphysik, D--69117 Heidelberg, Germany}
\affiliation{DESY, D--15738 Zeuthen, Germany}

\author{Michael K. Rosner\orcidB{}}
\affiliation{Max-Planck-Institut f\"ur Kernphysik, D--69117 Heidelberg, Germany}

\author{Charles Cheung}
\author{Sergey G. Porsev}
\author{Marianna S. Safronova}
\affiliation{Department of Physics and Astronomy, University of Delaware, Newark, Delaware 19716, USA}

\author{Steven Worm}
\affiliation{DESY, D--15738 Zeuthen, Germany}

\author{Dmitry Budker\orcidD{}}
\affiliation{Johannes Gutenberg-Universit{\"a}t Mainz, 55122 Mainz, Germany}
\affiliation{Helmholtz Institute Mainz, 55099 Mainz, Germany}
\affiliation{GSI Helmholtzzentrum für Schwerionenforschung GmbH, 64291 Darmstadt, Germany}
\affiliation{Department of Physics, University of California, Berkeley, CA 94720-7300, United States of America}

\author{Thomas Pfeifer\orcidL{}}
\affiliation{Max-Planck-Institut f\"ur Kernphysik, D--69117 Heidelberg, Germany}

\author{Jos\'e R. {Crespo L\'opez-Urrutia}\orcidJ{}}
\email[]{crespojr@mpi-hd.mpg.de}
\affiliation{Max-Planck-Institut f\"ur Kernphysik, D--69117 Heidelberg, Germany}

\author{Hendrik Bekker\orcidC{}}
\email[]{hbekker@uni-mainz.de}
\affiliation{Johannes Gutenberg-Universit{\"a}t Mainz, 55122 Mainz, Germany}
\affiliation{Helmholtz Institute Mainz, 55099 Mainz, Germany}
\affiliation{GSI Helmholtzzentrum für Schwerionenforschung GmbH, 64291 Darmstadt, Germany}

\date{\today}
	
\begin{abstract}
Optical transitions of highly charged ions can be very sensitive to hypothetical beyond-the-Standard-Model phenomena. Those near the $5s-4f$ level crossing, where the $5s$ and $4f$ are degenerate are especially promising. We present predictions from atomic theory and measurements of Os$^{15,16,17+}$ at an electron beam ion trap for identification of several transitions suitable for searches for a hypothetical fifth force and possible violations of local Lorentz invariance. The electric quadrupole (E2) transitions of Os$^{16+}$ that were found are especially suitable for frequency metrology due to their small linewidth of \SI{44}{\micro Hz}. Our calculations show the need for including enough inner-shell excitations to predict transition rates between configurations, which can otherwise be overestimated. Ultimately, the predicted interconfiguration transitions were too weak to be detected.
\end{abstract}
	
\maketitle

Very recently, by applying quantum logic \cite{schmidt_2005} to frequency metrology \cite{Diddams2000,Holzwarth2000} of sympathetically cooled  highly charged ions (HCI) \cite{schmöger_coulomb_2015}, a precision close to that of state-of-the-art optical clocks \cite{ludlow_optical_2015} for neutral and singly charged atoms has been reached at the German metrology institute Physikalisch-Technische Bundesanstalt (PTB) in Braunschweig \cite{micke_coherent_2020, king_2021,king_optical_2022,Cheung2025,chen_identification_2024,Wilzewski2025,Spiess2025}. This enables searches for physics beyond the Standard Model (BSM) \cite{Kozlov_2018} using, e.~g., King-plot analysis \cite{Berengut_2018,Rehbehn_2021,Rehbehn_2023} to probe hypothetical Yukawa interactions, as just demonstrated at PTB \cite{Wilzewski2025}. The sensitivity of HCI to various BSM effects, such as variation of the fine-structure constant $\alpha$, is highest near orbital crossings \cite{Berengut2010,berengut_electron-hole_2011} where the filling order  changes. However, predictions for the energy levels are extremely difficult in the interesting case of the $5s-4f$ orbital crossing. Here we investigate both theoretically and experimentally one of the most promising candidates for this crossing.
It was at first expected to provide laser-accessible optical transitions between fine-structure levels of the [Pd]$4f^{12} 5s^2$ and [Pd]$4f^{13} 5s$ configurations in the Nd-like iridium ($Z=77$) Ir$^{17+}$ ion. They were, however, not found, and recent predictions show that they should appear in the vacuum ultraviolet range instead~\cite{2020Ir}. In the present work, we study Nd-like Os$^{16+}$ ($Z=76$), which has a smaller splitting between the relevant configuration, with newer calculations predicting laser-accessible transitions there~\cite{Dzuba_2023}. Moreover, the lowest excited state couples to the ground state by an ultra-narrow electric quadrupole (E2) transition well suited for frequency metrology.

One central advantage of HCI is the strong binding of the outer electrons~\cite{Kozlov_2018}. Their wave function is thus less affected by external perturbations that hamper other systems. In contrast, its strong overlap with the nucleus enhances hypothetical BSM electron-neutron interactions accessible to the generalized King plot method~\cite{berengut_2020}. With its seven stable, naturally occurring isotopes, of which five have nuclear spin $I=0$, and its high nuclear charge and resulting relativistic effects, osmium is very well suited for such studies. In Os$^{16+}$, these effects are very different for $s$- and $f$-electrons, yielding for interconfiguration transitions an outstanding sensitivity to a potential $\alpha$ variation and to hypothetical violations of local Lorentz invariance~\cite{Dzuba_2023}.

\textit{Theory--} We perform large-scale configuration interaction (CI) calculations for Os$^{16+}$, treating the $[1s^2...4d^{10}]$ closed-shell core electrons and the 14 $f$ valence electrons with systematic accounting of correlation effects. The vast number of configurations with small individual weights requires the inclusion of many to achieve convergence. To support these demanding computations, we accelerate threefold the Hamiltonian construction using a bitstring-determinant storage and manipulation method developed for the \textsc{pCI} package~\cite{2025pCI} based on Ref.~\cite{Knowles84}. 

We construct our basis set based on previous work with Ir$^{17+}$~\cite{2020Ir} expanding it to include orbitals with principal quantum number up to $n=13$ for partial waves up to $l=7$, and comprising all orbitals up to $13g$, $12h$, $11i$, and $10k$. 
We gradually expand the basis set until convergence is reached for various correlation corrections, using a reference
30-electron CI calculation (that is, allowing excitations from the $5s$, $4f$, $4d$, and $4p$ shells) with a $7spdfg$ basis set.
We refer to allowing excitations from an inner shell as \textit{opening the shell}. 

First, we performed shell-by-shell CI calculations following Ref.~\cite{2020Ir} to account for inner-shell correlations. As in Ir$^{17+}$, we found the largest contributions from opening the $4d$ shell, shifting the configuration splitting by $\sim 8000$\,cm$^{-1}$, followed by opening the $4p$ and $4s$ shells, which give an overall contribution larger than the contribution from the innermost shells with $n \leq 3$. Although even-parity levels are largely unaffected, odd-parity levels shift by about $-2100$\,cm$^{-1}$. 
We then expand the basis set to $13g12h11i10k$ and extrapolate higher $n$ partial-wave contributions. 
Allowing excitations from $4f^{14}$ leads to negligible shifts for most states. 
Finally, we include (i) single and double excitations from $4f^{12}5p^2$ and single excitations from six dominant odd-parity configurations within 30-electron CI, and (ii) triple excitations from $4f^{12}5s^2$ and $4f^{14}$ within 14-electron CI. 

The results are shown in column ``\textsc{pCI}'' of Table~\ref{tab:energy_levels}. 
Details of these computations are described in the End Matter.
The largest energy shifts arise from including the inner shells with $n=3,4$, higher partial waves, and additional reference configurations.
The limitations of our local computing cluster restricted the full inclusion of these effects, potentially leading to a sizable underestimation.
We assign uncertainties accordingly to reflect these omissions.

\begin{table*}
\caption{\label{E1} Selection of E1 reduced matrix elements $\langle\,4f^{12}5s^2 \Vert D \Vert \,4f^{13}5s\rangle$ (in $10^{-3}$ atomic units, a.u.) and transition rates $A_{ab}$ (in s$^{-1}$) in different approximations. Calculations are performed with a $7spdf\!g$ basis set. The predicted wavelengths $\lambda$ (in nm) and their uncertainties are listed in the second column. The results obtained in the framework of 24-, 30-, and 60-electron CI are listed in columns ``$4d$ open,'' ``$4p$ open,'' and ``all open,'' respectively. The final values and uncertainties are listed under their respective columns, ``Final'' and ``Uncertainty'', and include the correlation of all 60 electrons, along with contributions from other correlation corrections. The transition rate uncertainties originate from those on $D$ and the energy uncertainties.}
 \begin{ruledtabular}
\begin{tabular}{lccccccccc} \multicolumn{1}{c}{Transition}& \multicolumn{1}{c}{$\lambda$ (nm)} &\multicolumn{5}{c}{$D$ ($\times10^{-3}$ a.u.)} & \multicolumn{3}{c}{$A_{ab}$ (s$^{-1}$)}\\ 
 \multicolumn{1}{c}{$a-b$}& \multicolumn{1}{c}{} & \multicolumn{1}{c}{$4d$ open}& \multicolumn{1}{c}{$4p$ open}& \multicolumn{1}{c}{all open} &\multicolumn{1}{c}{Final}& \multicolumn{1}{c}{Uncertainty}&\multicolumn{1}{c}{$4d$ open}&\multicolumn{1}{c}{Final}&\multicolumn{1}{c}{Uncertainty}\\
\hline \\[-0.7pc]
 $^3$F$_4 -\,  ^3$F$^\mathrm{o}_4$ &   451$^{+84}_{-61}$ & 1.93 & 0.31 & 0.37 & 0.12 & 0.27 & 12.62 &   0.04 & 0.16 \\[3pt]
 $^3$H$_4 -\,  ^3$F$^\mathrm{o}_4$ &   555$^{+135}_{-90}$ & 2.46 & 2.12 & 1.94 & 1.75 & 0.24 &  4.61 &   4.02 & 2.59 \\[3pt]
 $^3$H$_4 -\,  ^3$F$^\mathrm{o}_3$ &   743$^{+261}_{-153}$ & 3.61 & 2.45 & 2.34 & 2.12 & 0.21 &  3.34 &   2.47 & 1.98 \\[3pt]
 $^3$F$_4 -\,  ^3$F$^\mathrm{o}_3$ &   374$^{+56}_{-43}$ & 2.43 & 0.91 & 0.99 & 0.87 & 0.17 & 42.95 &   4.19 & 2.35 \\[3pt]
 $^3$P$_2 -\,  ^3$F$^\mathrm{o}_2$ &   426$^{+77}_{-56}$ & 1.48 & 1.19 & 1.11 & 1.01 & 0.11 & 10.61 &   5.36 & 2.74 \\[3pt]
 $^3$P$_2 -\,  ^1$F$^\mathrm{o}_3$ &   543$^{+132}_{-89}$ & 2.69 & 1.52 & 1.61 & 1.60 & 0.10 & 16.16 &   6.47 & 3.88 \\ 
\end{tabular}
   \end{ruledtabular}
\end{table*}

We calculated electric dipole (E1) reduced matrix elements $\langle 4f^{12}5s^2 \Vert D \Vert \,4f^{13}5s\rangle$ and found that the result strongly depends on which shells are opened.
In contrast, we find that reduced matrix elements of the magnetic dipole (M1) operator and the corresponding transition rates are only weakly affected by additional correlation corrections.
For the $4f^{12}5s^2\,^3$F$_4 -\, 4f^{13}5s\,^3$F$^\mathrm{o}_4$ E1 transition, our result at the reference $7spdf\!g$ level within the framework of the 24-electron CI (the column ``4d open'') agrees with the previous theory~\cite{Dzuba_2023}.
However, adding correlations from the $4p$ shell reduces the amplitude from $1.93\times10^{-3}$ a.u. to $0.12\times10^{-3}$ a.u., lowering the transition rate to below 1 s$^{-1}$, as shown in Table~\ref{E1}.
We attribute this effect to contributions from one-electron $4p-5s$ matrix elements.
We emphasize that calculations neglecting inner-shell correlations are inadequate and unreliable for guiding or interpreting experimental measurements.
Further suppression arises from the inclusion of the $h$ partial wave, likely due to the cumulative effect of many small but non-negligible admixtures.
Detection of these interconfiguration E1 transitions would uncover the long sought-after $5s-4f$ level crossing and reveal the wavelengths of the most $\alpha$-variation sensitive ones.
Table~\ref{E1} shows those with highest transition rates, which are therefore the most promising prospects for experimental observation.
Uncertainty values for the reduced matrix elements are assigned by adding the respective contributions in quadrature. For $^3$F$_4 -\,  ^3$F$^\mathrm{o}_4$, the uncertainty is greater than the value of the matrix element $D$, so $D$ should be taken as an order of magnitude estimate.

\textit{Measurements--} Osmium was investigated in an electron beam ion trap (EBIT)~\cite{levine_electron_1988,levine_use_1989,crespo_1999a}. The Heidelberg EBIT (HD-EBIT) ran with a 40~mA, \~400\,eV electron beam compressed by its $\SI{8}{T}$ magnetic field. This beam radially traps approximately $3\times10^{6}$ Os$^{16+}$ ions in its negative space-charge potential, which are axially confined by biased drift tubes, forming a cylindrical cloud of 50\,mm length and 0.1\,mm diameter. Osmium atoms are sourced from a tenuous molecular beam of osmocene, a volatile organometallic compound that the electron beam dissociates. Sequential electron-impact ionization and suppressed charge exchange in the 4-K environment allows selection of the desired charge state by setting the electron beam energy and axial trap depth. An image of the horizontal ion cloud was rotated and focused onto the vertical entrance slit of a $\SI{2}{m}$ focal-length grating spectrometer by a set of lenses and mirrors. For calibration, a diffuse reflector illuminated by a hollow-cathode lamp is moved in and out of an intermediate image plane between lenses. The spectrometer low-noise CCD camera uses exposure times of $\SI{1}{h}$ for the HCI. Dark images without trapped ions are recorded regularly to remove the stray light background and correct for the sensitivity of each camera pixel. Regular calibrations correct for thermal and mechanical drifts of the spectrometer. These standard procedures allowed us to reach wavelength fractional uncertainties as low as $\Delta \lambda/\lambda = 2.5 \cdot 10^{-7}$ \cite{Draganic2003,SoriaOrts2006,Windberger2015,Bekker2018,Bekker2019,Rehbehn_2021,Rehbehn_2023,Rosner_2024}.

In our initial observations, we detected a comprehensive list of Os$^{15,16,17+}$ lines, including very weak ones with a broadband $\SI{150}{grooves/mm}$ grating blazed for \SI{500}{nm} to cover the region 316--810\,$\SI{}{nm}$ and a $\SI{1800}{grooves/mm}$ holographic grating for the 227--$\SI{323}{nm}$ range. The broad entrance slit width of $\SI{150}{\micro m}$ yielded strong signals at reduced resolution. We set three different electron-beam energies for determining the charge states based on the line intensities. Next, lines assigned to Os$^{16+}$ and some belonging to Os$^{15+}$ and Os$^{17+}$ were re-measured at high resolving power using the $\SI{1800}{grooves/mm}$ or a $\SI{3600}{grooves/mm}$ grating with a $\SI{60}{\micro m}$ slit. In some cases, we measured their second diffraction order for better resolving power and to access more calibration lines. Each HCI line was recorded for at least 7 hours while up to 34 hours were spent on weak ones.

\begin{figure}
    \centering
    \includegraphics[width=\linewidth]{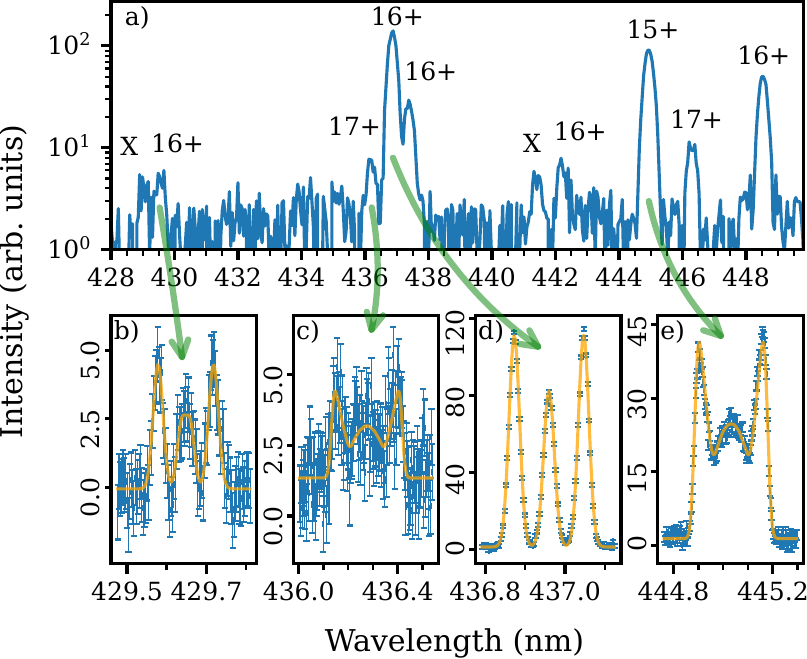}
    \caption{Examples of recorded spectra. a) Overview spectrum taken with the $\SI{150}{grooves/mm}$ grating showing lines of varying intensity of three charge states and two lines not originating from osmium (marked with an X). The bottom row shows spectra recorded at high resolving power with fits of our lineshape model in orange. b) The weakest identified transition belonging to Os$^{16+}$. c) Transition between the two lowest states of the Os$^{17+}$ $4f^{13}$ configuration. d) One of the strongest Os$^{16+}$ transitions. e) The ground state transition of Os$^{15+}$.}
    \label{fig:zeeman_fits}
\end{figure}

\begin{table}
    \centering
    \caption{Measured and inferred transition vacuum wavelengths $\lambda_\mathrm{vac}$ of Os$^{15+}$, Os$^{16+}$, and Os$^{17+}$ with identifications. Transition rates $A_{ab}$ are based on our \textsc{pCI} calculations for Os$^{16+}$ and AMBiT~\cite{Kahl_2019} ones for Os$^{15, 17+}$.}
    \begin{tabular}{llrrr}
        \hline
        \hline
        \vspace{-1.5ex} \\
        Config. & Transition & $\lambda_\mathrm{vac}$~(nm) & $A_{ab}$~(s$^{-1}$) & Type \\
        \vspace{-2ex} \\
        \hline
        \vspace{-2ex} \\
        
        Os$^{15+}$ & & & & \\
        \vspace{-2ex} \\
        $4f^{13} 5s^2$ & $^2$F$_{5/2}^\mathrm{o}$ - $^2$F$_{7/2}^\mathrm{o}$ & 445.03146(63) & 181 & M1 \\

        \vspace{-2ex} \\
        \hline
        \vspace{-2ex} \\
        
        Os$^{16+}$ & & & & \\
        \vspace{-2ex} \\
        $4f^{13} 5s$     & $^1$F$_3^\mathrm{o}$ - $^3$F$_4^\mathrm{o}$ & 360.11375(17) & 196 & M1 \\
         & $^1$F$_3^\mathrm{o}$ - $^3$F$_3^\mathrm{o}$ & 429.6482(11) & 6.5 & M1 \\
         & $^3$F$_2^\mathrm{o}$ - $^3$F$_3^\mathrm{o}$ & 547.7247(38) & 141 & M1 \\ 
        \vspace{-2ex} \\ 
        $4f^{12} 5s^2$ & $^3$P$_1$ - $^3$F$_2$ & 242.3348(17) & 91 & M1 \\
        &  & 242.34080(16)\,\textbf{R}  & &  \\
         & $^3$H$_4$ - $^3$F$_4$ & 247.1159(8) & 13 & M1 \\
        &  & 247.11704(19)\,\textbf{R}  & &  \\ 
         & $^3$H$_4$ - $^3$H$_5$ & 356.28870(9) & 249 & M1 \\
         & $^3$D$_2$ - $^3$F$_2$ & 395.74681(36) & 136 & M1 \\
         & $^3$F$_3$ - $^3$F$_4$ & 436.95983(16) & 189 & M1 \\
         & $^3$P$_2$ - $^3$D$_2$ & 437.48246(36) & 165 & M1 \\
         & $^3$D$_2$ - $^3$F$_3$ & 448.62855(31) & 129 & M1 \\
         & $^3$H$_5$ - $^3$H$_6$ & 472.19841(26) & 248 & M1 \\
         & $^1$G$_4$ - $^3$F$_4$ & 491.79063(29) & 92 & M1 \\ 
         & $^3$H$_4$ - $^1$G$_4$ & 496.70195(72) & 77 & M1 \\
         & $^3$H$_4$ - $^3$F$_3$ & 568.7831(25)  & 8.1 & M1 \\ 
         &  & 568.7876(11)\,\textbf{R}  &  &  \\
         & $^3$P$_1$ - $^3$D$_2$ & 625.17496(60) & 43 & M1 \\
        \vspace{-2ex} \\
         & $^3$F$_2$ - $^3$F$_4$ & 502.3391(12)\,\textbf{R} & $3.99\times10^{-3}$ & E2 \\ 
         & $^3$F$_4$ - $^3$H$_6$ & 1139.2103(78)\,\textbf{R} & $4.36\times10^{-5}$ & E2 \\ 

        \vspace{-2ex} \\
        \hline
        \vspace{-2ex} \\
        
        Os$^{17+}$ & & & & \\
        $4f^{13}$       & $^2$F$_{5/2}^\mathrm{o}$ - $^2$F$_{7/2}^\mathrm{o}$ & 436.2751(20) & 185 & M1 \\
        $4f^{12} 5s^1$ & $^3$H$_{11/2}$ - $^4$H$_{13/2}$ & 363.986(21) & 296 & M1 \\
        $4f^{11} 5s^2$ & $^4$I$_{13/2}^\mathrm{o}$ - $^4$I$_{15/2}^\mathrm{o}$ & 500.689(25) & 319 & M1 \\

        \vspace{-2ex} \\
        \hline
        \hline
        \vspace{-2ex} \\
        \multicolumn{3}{l}{\textbf{R}: Ritz-Rydberg combination} & & \\
    \end{tabular}
    \label{tab:measurement_results}
\end{table}

\begin{figure}
    \centering
    \includegraphics[width=\linewidth]{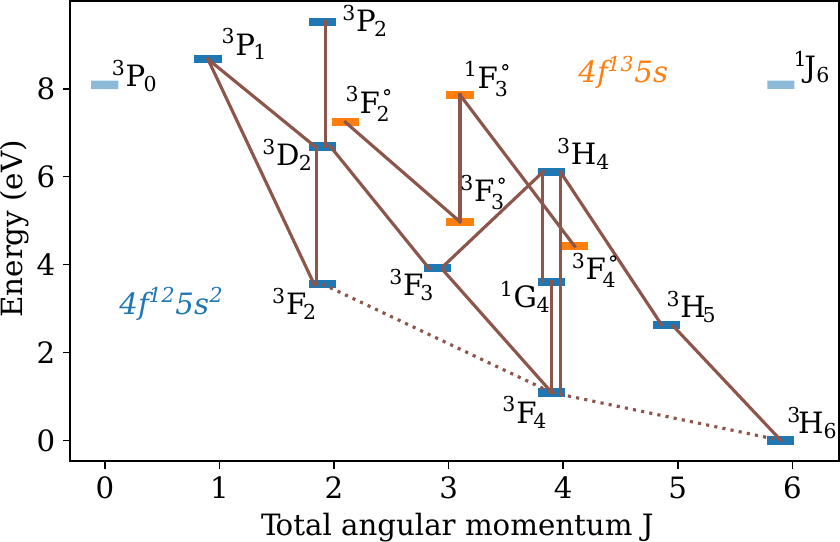}
    \caption{Level scheme of Os$^{16+}$ based mainly on measurements, c.f. Table~\ref{tab:energy_levels}. The measured M1 transitions are shown as solid lines and the E2 clock transitions are dashed. Disconnected level energies and the interconfiguration splitting are based on our \textsc{pCI} predictions.
    }
    \label{fig:grotrian}
\end{figure}

\textit{Identifications--} We assigned lines to Os$^{16+}$ by comparison with theory transition energies and by studying their characteristic Zeeman splitting in the $\SI{8}{T}$ field of HD-EBIT.
We modeled the line shape based on the total angular moment $J$ and the predicted $g_J$ factors (see Table~\ref{tab:energy_levels}).
The hyperfine structure of the $^{187,189}$Os isotopes was not resolvable.
Relative amplitudes of the Zeeman components were derived from the Clebsch-Gordan coefficients of the $|J, m_J\rangle$ states and accounting for the multipole-radiation patterns.
Fits of this model allowed us to unambiguously identify 15 M1 transitions within configurations (see Figure~\ref{fig:zeeman_fits} and~Table~\ref{tab:measurement_results}).
Several identifications are supported by previous analysis of wavelength scaling laws versus nuclear charge~\cite{Windberger2015}.
The splitting between the $^3$P$_1$ - $^3$F$_2$ states and $^3$H$_4$ - $^3$F$_3$ ones was determined by direct measurement of transitions between them, but also by measurement of transitions connecting to an intermediate state as can be seen in the level scheme in Fig.~\ref{fig:grotrian}.
These so-called Ritz-Rydberg combinations are further evidence supporting the correctness of identifications.

Our M1 line identifications fully reconstruct the fine structure of the Os$^{16+}$ $4f^{13} 5s$ configuration and nearly the full $4f^{12} 5s^{2}$ one, with levels listed in Table\,\ref{tab:energy_levels} and displayed in Fig.\,\ref{fig:grotrian}.
Our most advanced \textsc{pCI} calculations agree best out of all theory predictions with an average deviation of only 0.6~\% compared to the experimental values.
Based on the experimentally determined level energies, we obtained the energies of the ultra-narrow E2 transitions from the $^{3}$H$_{6}$ ground state to the first excited state $^{3}$F$_{4}$ with an uncertainty of 9.3 GHz and from there to the $^{3}$F$_{2}$ state with 584 MHz uncertainty (see Table~\ref{tab:measurement_results}).
This narrows the search range for precision laser spectroscopy, and should reduce the time needed for it with the methods recently developed at PTB~\cite{ChenPRR2024}.
Subsequent quantum logic spectroscopy on these E2 transitions, possibly by using auxiliary M1 ones, has the potential to result in the most accurate optical atomic clock to date with highest sensitivity to local Lorentz invariance~\cite{king_optical_2022, Dzuba_2023}.

\begin{table*}
    \centering
    \caption{\label{tab:energy_levels} Level energies based on  measured transitions (see Table\,\ref{tab:measurement_results}) with fitted $g$-factors, compared with calculations from Ref.~\cite{Dzuba_2023} and from this work using \textsc{ambit}~\cite{Kahl_2019}, \textsc{fac}~\cite{gu_2008}, and \textsc{pCI}.}
    \begin{tabular}{llllrrrrrrr}
        \hline
        \hline
        \vspace{-1.5ex} \\
        \multicolumn{2}{c}{} & \multicolumn{2}{c}{Measured} & \multicolumn{2}{c}{Ref.~\cite{Dzuba_2023}} & \multicolumn{2}{c}{\textsc{ambit}} & \multicolumn{1}{c}{\textsc{fac}} & \multicolumn{2}{c}{\textsc{pCI}} \\
        
        Config. & Level & $E$~(eV) & $g$-factor & $E$~(eV) & $g$-factor & $E$~(eV) & $g$-factor & $E$~(eV) & $E$~(eV) & $g$-factor \\
        \vspace{-2ex} \\
        \hline
        \vspace{-2ex} \\
        $4f^{12}5s^2$ & $^3$H$_6$ & 0 & 1.147(2) & 0 & 1.164 & 0 & 1.164 & 0 & 0 & 1.164 \\
         & $^3$F$_4$ & 1.0883346(74) & 1.137(5) & 1.188 & 1.137 & 1.133 & 1.138 & 1.235 & 1.101(10) & 1.142 \\
         & $^3$H$_5$ & 2.6256801(14) & 1.017(3) & 2.781 & 1.033 & 2.672 & 1.033 & 2.618 & 2.611(16) & 1.033 \\
         & $^3$F$_2$ & 3.556472(13) & 0.99(6) & 3.803 & 0.824 & 3.724 & 0.834 & 3.950 & 3.585(61) & 0.837 \\
         & $^1$G$_4$ & 3.6094115(59) & 0.986(5) & 3.799 & 0.989 & 3.690 & 0.992 & 3.688 & 3.602(17) & 0.989\\
         & $^3$F$_3$ & 3.9257624(84) & 1.060(5) & 4.157 & 1.083 & 4.057 & 1.083 & 4.128 & 3.928(36) & 1.083 \\
         & $^3$H$_4$ & 6.1055603(23) & 0.904(5) & 6.210 & 0.927 & 6.215 & 0.920 & 6.155 & 6.085(37) & 0.920 \\
         & $^3$D$_2$ & 6.689389(10) & 1.119(4) & 7.196 & 1.130 & 6.998 & 1.124 & 7.311 & 6.734(119) & 1.129 \\
         & $^1$J$_6$ & - & - & - & - & 8.325 & 1.0026 & 8.579 & 8.091(169) & 1.003 \\ 
         & $^3$P$_0$ & - & - & - & - & 8.449 & 0 & 8.980 & 8.088(186) & 0 \\ 
         & $^3$P$_1$ & 8.672581(12) & 1.469(12) & - & - & 9.117 & 1.500 & 9.565 & 8.745(192) & 1.500 \\
         & $^3$P$_2$ & 9.523428(13) & 1.188(4) & - & - & 9.916 & 1.209 & 10.257 & 9.565(138) & 1.201 \\
        \vspace{-2ex} \\
        $4f^{13}5s$ & $^3$F$_4^\mathrm{o}$ & x & 1.254(4) & 4.080 & 1.250 & 3.702 & 1.250 & 3.416 & 3.853(433) & 1.250 \\
         & $^3$F$_3^\mathrm{o}$ & x+0.5572030(88) & 1.042(4) & 4.641 & 1.0524 & 4.297 & 1.054 & 4.006 & 4.416(433) & 1.053 \\
         & $^3$F$_2^\mathrm{o}$ & x+2.820826(24) & 0.663(4) & - & - & 6.562 & 0.667 & 6.147 & 6.652(423) & 0.667 \\
         & $^1$F$_3^\mathrm{o}$ & x+3.4429177(15) & 1.043(5) & - & - & 7.226 & 1.029 & 6.830 & 7.283(425) & 1.030 \\
        \vspace{-2ex} \\
        $4f^{14}$ & $^1$S$_0$ & - & 0 & - & - & 14.155 & 0 & 14.491 & 12.786 & 0 \\
        \vspace{-1.5ex} \\
        \hline
        \hline
    \end{tabular}
\end{table*}

Having identified all strong lines as M1 transitions, the remaining weak ones could be due to the strongest E1 interconfiguration transitions, such as the ones listed in Table~\ref{E1}.
However, transition rates alone do not determine the observed line intensities; the population in the initial state and the detection efficiency at the transition wavelength must also be considered.
Owing to the identification of the $^{3}$H$_{4}$ - $^{3}$F$_{3}$ transition, we know that its initial state is sufficiently populated to detect a transition with a rate of 6.5~s$^{-1}$ with a signal-to-noise ratio (SNR) of 5.
Therefore, the E1 transitions from $^{3}$H$_{4}$ to $^{3}$F$_{4}^\mathrm{o}$ is expected to have an SNR of only slightly above 1 and the one to $^{3}$F$_{3}^\mathrm{o}$ is expected to be too weak to observe due to the reduced detection efficiency far from the grating optimum.
We find several unidentified candidates for the latter transition in the recorded spectra, but either their line shapes did not match the predicted one or were too weak to measure accurately.
These candidates could alternatively be M1 transitions between highly excited metastable states~\cite{Rosner_2024}.
They can also originate from tiny impurities in the sample of HCI which, despite careful adjustment of the trapping potential and regular reloading of an osmium sample, can originate from the EBIT's barium-tungsten based electron emitter.

To predict intensities of other E1 lines we rely on level-population predictions obtained using the collisional radiative modeling (CRM) module of the Flexible Atomic Code~\cite{gu_2008}.
Our model, previously used to accurately predict line intensities in the extreme ultraviolet spectrum of Os$^{16+}$, shows that the steady-state populations of the studied states can differ by more than an order of magnitude~\cite{BekkerEUV2015}.
We confirm the accuracy of these predictions in the optical range by comparing the intensity of two simultaneously measured, closely spaced lines at \SI{437}{nm}, where differences in detection efficiency are negligible.
CRM calculations yield a population ratio in the upper states $P(^3\mathrm{F}_3)/P(^3\mathrm{P}_2) = 6.5$, which together with the predictions from Table~\ref{tab:measurement_results}, leads to a predicted ratio of line intensities of 5.7, close to the measured value of 5.3.
Combining these results with the predicted E1 rates and detector efficiency shows that the lines are too weak to be definitively detected with the employed setup,

Several strong M1 lines of Os$^{15+}$ and Os$^{17+}$ could also be identified using the described methods (see Table\,\ref{tab:measurement_results}).
In these cases, energy levels with an accuracy of approximately 2~\%, $g_J$ factors, and transition rates were also calculated using AMBiT~\cite{Kahl_2019}.
Not all transitions were measured at high resolving power, but the resulting wavelength uncertainties are still sufficiently small to perform laser spectroscopy on them in a reasonable time with newly developed techniques~\cite{chen_identification_2024}.
The provided catalog of laser-addressable transitions paves the way for generalized King-plot searches of a hypothetical fifth force when several transitions are measured in different isotopes~\cite{berengut_2020, Wilzewski2025}.

\textit{Conclusions--} We measured a large set of forbidden optical M1 transitions in Os$^{15+,16+,17+}$ ions and experimentally determined most level energies of the Os$^{16+}$ $4f^{12}5^2$ and $4f^{13}5s$ level-crossing configurations.
These are in excellent agreement with our theory, which furthermore predicts the interconfiguration E1 transition rates to be much lower than previously thought due to insufficient consideration of inner-shell excitations.
Consequently, interconfiguration transitions remained undetectable, notwithstanding the high sensitivity of our experimental setup, which enables the observation and accurate wavelength determination of weak M1 optical transitions through spontaneous emission in the laboratory. In principle, longer exposure times and background reduction could allow detection of such interconfiguration transitions if there are no fast competing transitions depopulating the upper state.
Nonetheless, we discovered two ultra-narrow clock transitions in Os$^{16+}$ suitable for frequency metrology and searches for local Lorentz invariance.
Precision laser spectroscopy of the reported ground-state transitions in three charge states and seven stable osmium isotopes is expected to yield excellent sensitivity to a hypothetical fifth force.

\section*{Acknowledgements}
HB and JRCLU thank Julian Berengut for helpful discussions.
The theoretical work has been supported in part by the US NSF Grant  No. PHY-2309254, US Office of Naval Research Grant N000142512105  and by the European Research Council (ERC) under the Horizon 2020 Research and Innovation Program of the European Union (Grant Agreement No. 856415).
The calculations in this work were done through the use of Information Technologies resources at the University of Delaware, specifically the high-performance Caviness and DARWIN computer clusters.
HB was supported in part by the project ``Quantum Sensing for Fundamental Physics (QS4Physics)’’ funded by the Innovation pool of the research field Helmholtz Matter of the Helmholtz Association.

\bibliography{literature.bib}

\clearpage
\onecolumngrid
\section*{End Matter}
\twocolumngrid

\textit{Details of computations--} We treat Os$^{16+}$ as a 14-valence electron system with a [$1s^2,\dots,4d^{10}$] closed core. We start by solving the Dirac-Hartree-Fock equations for the $[1s^2...4d^{10}]\,4f^{13}5s$ configuration to construct one-electron basis orbitals in the central field approximation. Then, all electrons are frozen, and an electron is promoted from the $5s$ to the $5p$ and $5d$ shells to construct the corresponding orbitals. The $5g$, $6h$, $7i$, and $8k$ orbitals are constructed from the $4f$ orbital, while the remaining virtual orbitals are formed using a recurrent procedure described in Refs.~\cite{KozPorFla96,KozPorSaf15}. All orbitals are built on a radial grid within a compact spherical cavity of $R=5$ a.u., which improves convergence to the principal quantum number $n$. Both the Coulomb and Breit interactions are included throughout. 

The CI many-electron wave function is expressed as a linear combination of all distinct states with a given total angular momentum $J$ and parity: $\Psi_J=\sum_ic_i\Phi_i$. The corresponding energies and wave functions are obtained by solving the time-independent relativistic many-electron Schr\"odinger equation $H_n\Psi_n=E_n\Psi_n$.
Because a large fraction of configurations ($\gtrsim\!80\%$) contribute significantly even with modest basis sets, subselection methods~\cite{2025pCI,2025NNpCI} could not be used to optimize the CI space efficiently to reduce the computational workload. 

The energies obtained from CI are listed in Table~\ref{summary}, along with uncertainties calculated by adding all correlation contributions in quadrature. We set a baseline by allowing all single and double (SD) excitations from the $4f^{12}5s^2$ and $4f^{13}5s$ reference configurations to the $7spdf\!g$ basis set. We then performed CI calculations with basis sets of incrementally higher principal quantum number $n$ until either the energy converges or until the computation is too large, then repeat this procedure for higher partial waves. In total, we include all orbitals up to $13spdf\!g12h11i10k$. 
To handle computations that were too large, we estimated the $n\rightarrow\infty$ contributions for $l > 5$ partial waves by extrapolating their respective energy convergence. The resulting values are listed in column ``extrap.''

We performed shell-by-shell CI in a small $7spdf\!g$ basis set to obtain contributions of electron correlations of the inner shells following Ref.~\cite{2020Ir} and list the results in Table~\ref{shellCI}. 
We begin with a base computation allowing excitations only from the $5s$ and $4f$ orbitals, then systematically obtain contributions from opening the inner shells until all 60 electrons are correlated. 
We find that the largest effect to the configuration splitting comes from the inclusion of the inner $n\geq3$ electron shells into the CI. 

We also examined how energy levels change when excitations from additional reference configurations are included -- a computationally demanding task due to the vast number of resulting configurations. Specifically, we evaluated the effect of including the $4f^{14}$ configuration, which incorporates both the shell-by-shell CI expansion and an extended $13spdf\!g7h$ basis. These results are listed in column ``$+4f^{14}$'' of Table~\ref{summary}. Most levels show small shifts (under 20 cm$^{-1}$), except for the $4f^{12}5s^2\,^3$P$_0$ state, which decreases by 499 cm$^{-1}$. 

In the ``extras'' column, we list contributions from two additional sources: (i) SD excitations from the $4f^{12}5p^2$ even-parity configuration and single excitations from six highly-weighted odd-parity configurations in a 30-electron CI with the $4p$ shell open and (ii) triple excitations from $4f^{12}5s^2$ and $4f^{14}$ in a 14-electron CI. 
We limited the odd-parity excitations to singles because including SD excitations from the second most contributing odd-parity configuration, $4f^{11} 5s 5f^2$, was computationally infeasible. 
These additional reference configurations produced small differences (below 21 cm$^{-1}$) for $4f^{12}5s^2$, but led to significant shifts for $4f^{14}\,^1$S$_0$ (1385 cm$^{-1}$) and $4f^{13}5s$ levels ($\sim 920$ cm$^{-1}$). 
Triple excitations contributed modestly to $4f^{14}\,^1$S$_0$ (588 cm$^{-1}$) and $4f^{12}5s^2$ ($-156$ cm$^{-1}$), but significantly to $4f^{13}5s$ ($\sim 1500$ cm$^{-1}$).

\begin{table*} [htb]
\label{tab:calculated_levels}
\caption{\label{summary} Contributions to Os$^{16+}$ energies (in cm$^{-1}$) obtained using CI with different correlation corrections. The base run is performed with a small $7spdf\!g$ basis set with the $4p$ shell open. The contribution from correlating all 60 electrons in the framework of 60-electron CI is given in the column labeled ``$+$all open.'' Contributions from allowing excitations to higher basis orbitals are given in the respective columns labeled ``+$nl$''. Contributions from including $4f^{14}$ and others as reference configurations are given in the columns ``$+4f^{14}$'' and ``+extras,'' respectively. The extrapolation of missing contributions from the basis set expansion is given in the column ``+extrap.'' The last two columns display the final energies and uncertainties.}
 \begin{ruledtabular}
\begin{tabular}{lcccccccccccc} 
\multicolumn{2}{c}{}&
\multicolumn{1}{c}{$7g$}&\multicolumn{1}{c}{$7g$}& \multicolumn{1}{c}{}&
\multicolumn{1}{c}{}& \multicolumn{1}{c}{}&  \multicolumn{1}{c}{}&
\multicolumn{1}{c}{}&\multicolumn{1}{c}{}&\multicolumn{1}{c}{}& \multicolumn{1}{c}{} & \multicolumn{1}{c}{}\\
\multicolumn{2}{c}{Configuration}&
\multicolumn{1}{c}{$4p$ open}&\multicolumn{1}{c}{$+$all open}& \multicolumn{1}{c}{$+$(8-13)$g$}&
\multicolumn{1}{c}{$+$(6-12)$h$}& \multicolumn{1}{c}{$+$(7-11)$i$}&  \multicolumn{1}{c}{$+$(8-10)$k$}&
\multicolumn{1}{c}{$+4f^{14}$}&\multicolumn{1}{c}{$+$extras}&\multicolumn{1}{c}{$+$extrap.}& \multicolumn{1}{c}{Final} & \multicolumn{1}{c}{Uncertainty}\\
\hline \\[-0.7pc]
$4f^{12} 5s^2$ & $^3$H$_6$   &      0 &     0 &     0  &     0  &    0  &    0  &      0 &    0 &    0  &     0 &    0 \\
               & $^3$F$_4$   &   9067 &   -28 &   -35  &   -55  &  -40  &   -9  &     -8 &    7 &  -16  &  8883 &   80 \\
               & $^3$H$_5$   &  20970 &   120 &   -15  &    29  &   -3  &   -1  &      0 &  -40 &   -1  & 21061 &  130 \\
               & $^3$F$_2$   &  29825 &  -212 &   -84  &  -423  &  -96  &  -20  &    -12 &  -40 &  -22  & 29050 &  490 \\
               & $^1$G$_4$   &  29055 &   122 &   -40  &   -18  &  -11  &   -3  &     -4 &  -44 &  -10  & 28913 &  140 \\
               & $^3$F$_3$   &  32028 &    44 &   -43  &  -272  &   -9  &   -3  &     -6 &  -58 &   -3  & 31679 &  290 \\
               & $^3$H$_4$   &  48881 &   265 &   -62  &   114  &  -22  &   -6  &     -5 &  -66 &  -20  & 49079 &  300 \\
               & $^3$D$_2$   &  55680 &  -161 &  -112  &  -931  &  -32  &  -10  &    -14 & -107 &    2  & 54314 &  960 \\
               & $^1$I$_6$   &  66963 &   -20 &  -121  & -1352  &  -46  &   -8  &    -23 &  -79 &  -57  & 65255 & 1360 \\
               & $^3$P$_0$   &  67635 &  -199 &  -160  & -1393  &  -23  &  -10  &   -520 &  -83 &  -13  & 65234 & 1500 \\ 
               & $^3$P$_1$   &  72541 &  -165 &  -129  & -1531  &    7  &   -2  &    -16 & -164 &   -4  & 70537 & 1550 \\ 
               & $^3$P$_2$   &  78771 &  -182 &  -157  & -1072  &  -22  &   -5  &    -17 & -147 &  -22  & 77147 & 1110 \\ [0.5pc]
$4f^{13}5s$    & $^3$F$^\mathrm{o}_4$ &  33684 & -2195 &  -102  & -1234  & -622  &   21  &     20 & 2220 & -717  & 31076 & 3490 \\
               & $^3$F$^\mathrm{o}_3$ &  38288 & -2173 &  -106  & -1274  & -640  &   15  &     20 & 2210 & -721  & 35619 & 3490 \\
               & $^3$F$^\mathrm{o}_2$ &  56181 & -2083 &  -121  & -1234  & -620  &   20  &     20 & 2201 & -716  & 53649 & 3410 \\
               & $^1$F$^\mathrm{o}_3$ &  61408 & -2072 &  -131  & -1307  & -652  &   10  &     20 & 2195 & -728  & 58744 & 3430 \\                                                                                                                      
\end{tabular}
\end{ruledtabular}
\end{table*}

\begin{table*} [htb]
\caption{\label{shellCI} Contributions to Os$^{16+}$ energies (in cm$^{-1}$) obtained using CI with excitations from the increasing number of atomic subshells. The results of the base run, performed with a small $7spdf\!g$ basis set with the $5s$ and $4f$ shells open, are listed under column `$5s4f$`. Contributions from opening subshells are given separately in the columns labeled ``$nl$ contr.'', where $nl$ indicates the subshell opened. The final energies with all 60 electrons correlated by CI are listed in column ``Final.''}
 \begin{ruledtabular}
\begin{tabular}{lcccccccccccc} \multicolumn{2}{c}{Configuration}&
  \multicolumn{1}{c}{$5s4f$}&\multicolumn{1}{c}{$4d$}& \multicolumn{1}{c}{$4p$}&
\multicolumn{1}{c}{$4s$}& \multicolumn{1}{c}{$3d$}&  \multicolumn{1}{c}{$3p$}&
\multicolumn{1}{c}{$3s$}&\multicolumn{1}{c}{$2p$}& \multicolumn{1}{c}{$2s$} &\multicolumn{1}{c}{$1s$}& \multicolumn{1}{c}{Final}\\
\multicolumn{3}{c}{}&
  \multicolumn{1}{c}{contr.} &\multicolumn{1}{c}{contr.} &\multicolumn{1}{c}{contr.} &\multicolumn{1}{c}{contr.} &\multicolumn{1}{c}{contr.} &\multicolumn{1}{c}{contr.} &\multicolumn{1}{c}{contr.} &\multicolumn{1}{c}{contr.}& \multicolumn{1}{c}{contr.} &\multicolumn{1}{c}{}\\
\hline \\[-0.7pc]
$4f^{12} 5s^2$& $^3$H$_6$   &     0 &     0 &    0 &     0  &   0  &    0 &    0 &   0 &   0 &  0 &      0 \\
              & $^3$F$_4$   &  9630 &  -113 & -450 &    10  &  -1  &  -40 &    1 &   1 &   0 &  0 &   9039 \\
              & $^3$H$_5$   & 21060 &   -83 &   -8 &     9  &  76  &   37 &   -2 &   1 &   0 &  0 &  21090 \\
              & $^3$F$_2$   & 29496 &  -166 & -275 &    18  &  84  &   20 &   -1 &   2 &   0 &  0 &  29177 \\
              & $^1$G$_4$   & 30895 &  -459 & -611 &  -145  &  33  &  -91 &  -13 &   3 &   1 &  0 &  29614 \\
              & $^3$F$_3$   & 32655 &  -317 & -310 &   -12  &  75  &  -17 &   -5 &   2 &   0 &  0 &  32072 \\
              & $^3$H$_4$   & 49477 &  -210 & -386 &    46  & 158  &   60 &   -1 &   2 &   0 &  0 &  49145 \\
              & $^3$D$_2$   & 57275 & -1004 & -591 &  -125  &  80  & -107 &  -15 &   5 &   1 &  0 &  55519 \\
              & $^1$I$_6$   & 68808 & -1647 & -198 &   -40  &  94  &  -68 &  -10 &   3 &   1 &  0 &  66943 \\
              & $^3$P$_0$   & 69355 & -1105 & -615 &   -94  &  80  & -182 &  -16 &  10 &   2 &  0 &  67436 \\ 
              & $^3$P$_1$   & 74623 & -1575 & -507 &   -82  &  83  & -159 &  -16 &   8 &   2 &  0 &  72377 \\ 
              & $^3$P$_2$   & 80751 & -1192 & -788 &  -197  & 140  & -110 &  -23 &   7 &   1 &  0 &  78589 \\ [0.5pc]
$4f^{13}5s$   & $^3$F$^\mathrm{o}_4$ & 26041 &  8200 & -557 & -1824  & 333  & -250 & -323 & -81 & -47 & -4 &  31489 \\
              & $^3$F$^\mathrm{o}_3$ & 30599 &  8227 & -539 & -1811  & 342  & -247 & -326 & -81 & -47 & -4 &  36115 \\
              & $^3$F$^\mathrm{o}_2$ & 48599 &  8127 & -545 & -1812  & 405  & -218 & -325 & -81 & -47 & -4 &  54098 \\
              & $^1$F$^\mathrm{o}_3$ & 53597 &  8318 & -507 & -1792  & 404  & -220 & -331 & -81 & -47 & -4 &  59337 \\             
\end{tabular}
\end{ruledtabular}
\end{table*}

\end{document}